\documentstyle[aps, manuscript]{revtex}
\tightenlines
\draft
\begin{document}
\title{Metric fluctuations and its evolution during inflation}
\author{$^1$Mariano Anabitarte\footnote{
E-mail address: anabitar@mdp.edu.ar}
and $^{1,2}$Mauricio Bellini\footnote{
E-mail address: mbellini@mdp.edu.ar}}
\address{$^1$Departamento de F\'{\i}sica, Facultad de Ciencias
Exactas y Naturales
Universidad Nacional de Mar del Plata,
Funes 3350, (7600) Mar del Plata, Buenos Aires, Argentina.\\
and \\
$^2$ Consejo Nacional de Investigaciones Cient\'{\i}ficas y
T\'ecnicas (CONICET)}

\vskip .5cm
\maketitle
\begin{abstract}
We discuss the evolution of the fluctuations in a
symmetric $\phi_c$-exponential potential which provides a power-law
expansion during inflation using
both, the gauge invariant field $\Phi$ and the Sasaki-Mukhanov field.
\end{abstract}
\vskip .2cm                             
\noindent
Pacs numbers: 98.80.Cq \\
\vskip 1cm
\section{Introduction and motivation}

It is now widely accepted that dominant cause of structure in the
Universe is a spatial perturbation. This perturbation is present
on cosmological scales a few Hubble times before these scales enter
the horizon, at which stage it is time-independent with an almost
flat spectrum. One of the main objectives of theoretical cosmology
is to understand its origin\cite{Guth}.
The usual assumption is that the curvature
perturbation originates during inflation of the slow-rolling inflaton
field. As cosmological scales leave the horizon, the quantum fluctuation
is converted to a classical gaussian perturbation with an almost
flat spectrum, generating inmediately the required curvature perturbation
which is constant until the approach of horizon entry\cite{1}. This idea has
the adventage that prediction for the spectrum is independent
of what goes on between the end of inflation and horizon entry.
The spectrum depends only on the form of the potential and on the theory
of gravity during inflation, provinding, therefore, a direct probe
of conditions during this era.

Stochastic inflation has played an important role in inflationary cosmology
in the last two decades.
It
proposes to describe the dynamics of
this quantum field on the basis of two pieces: the homogeneous
and inhomogeneous
components\cite{starobinsky,sasaki,yi,hm,habib,mijic,copeland,BCMS}.
Usually the homogeneous one is interpreted
as a classical field $\phi_c(t) $
that arises from the vacuum expectation value of
the quantum field. The inhomogeneous component $\phi(\vec x,t)$
are the quantum fluctuations.
The field that take into account only the modes with wavelengths
larger than the now observable universe is called coarse-grained field
and its dynamics is described by a second order
stochastic equation\cite{BCMS,riotto}.
Since these perturbations are classical on super Hubble scales,
in this sector one can make a standard stochastic
treatment for the coarse-grained matter field\cite{BCMS}. The IR sector is
very important because the spatially inhomogeneities in
super Hubble inflationary scales would explain the present day
observed matter structure in the universe.

In this work we consider gauge-invariant fluctuations of the metric
in the early inflationary universe\cite{1}. Metric fluctuations
are here considered in the framework of the linear perturbative corrections.
The scalar metric perturbations are spin-zero projections of the graviton,
which only exists in nonvacuum cosmologies. The issue of gauge invariance
becomes critical when we attempt to analyze how the scalar metric perturbations
produced in the early universe influence a background globally flat isotropic
and homogeneous universe. This allows us to formulate the problem of the
amplitude for the scalar metric perturbations on the evolution of the
background Friedmann-Robertson-Walker (FRW) universe in a coordinate-independent
manner at every moment in time. On the other hand, the Sasaki-Mukhanov (SM)
field takes into account both, metric and inflaton fluctuations\cite{sm}.
One of the aims of this work is the study of the evolution of the SM field
during inflation
to make a comparison with gauge-invariant metric fluctuations.

\section{Fluctuations}

Matter field fluctuations are responsible for metric fluctuations
around the background FRW metric. When
these metric fluctuations do not depend
on the gauge, the perturbed globally flat
isotropic and homogeneous universe is described by\cite{1}
\begin{equation}\label{m}
ds^2 = (1+2\psi) \  dt^2 - a^2(t) (1-2\Phi) \  dx^2,
\end{equation}
where $a$ is the scale factor of the universe and ($\psi$, $\Phi$) are
the gauge-invariant perturbations of the metric.
In the particular
case where the tensor $T_{\alpha\beta}$ is diagonal, one obtains:
$\Phi = \psi$\cite{1}.
We consider a semiclassical expansion for the
inflaton field $\varphi(\vec x,t) = \phi_c(t) + \phi(\vec x,t)$\cite{BCMS},
with expectation values
$\left<0|\varphi|0\right> = \phi_c(t)$ and $\left<0|\phi|0\right>=0$. Here,
$\left.|0\right>$ is the vacuum state.
Due to $\left<0|\Phi|0\right> =0$, the expectation
value of the metric (\ref{m}) gives the background
metric that describes a flat FRW spacetime: $\left<ds^2\right>
= dt^2-a^2 dx^2$.

After linearizing the Einstein equations in terms of $\phi$ and $\Phi$,
one obtains
\begin{eqnarray}
\ddot\Phi &+& \left(H
- 2 \frac{\ddot\phi_c}{\dot\phi_c} \right)
\dot \Phi - \frac{1}{a^2} \nabla^2 \Phi +2\left(
\dot H - H \frac{\ddot\phi_c}{\dot\phi_c}\right) \Phi =0, \label{1}\\
\frac{1}{a}& \frac{d}{dt}& \left( a \Phi \right)_{,\beta} = 
\frac{4\pi}{M^2_p} \left(\dot\phi_c \phi\right)_{,\beta} , \\
\ddot\phi& +& 3 H \dot\phi -
\frac{1}{a^2} \nabla^2 \phi + V''(\phi_c) \phi 
+ 2 V'(\phi_c) \Phi- 4 \dot\phi_c \dot\Phi =0, \label{A}
\end{eqnarray}
where $\beta = 0,1,2,3$, $a$ is the scale factor of the universe
and the prime denotes the derivative with respect to
$\phi_c$. The dynamics of $\phi_c$ is given by the equations
\begin{equation}\label{dyn}
\ddot\phi_c + 3 H\dot\phi_c + V'(\phi_c)=0,
\qquad \dot\phi_c = -\frac{M^2_p}{ 4\pi} H',
\end{equation}
and $H=\dot a/a$ is the Hubble parameter.
Furthermore, the scalar potential can be written in terms of
the Hubble parameter
\begin{equation}
V(\phi_c) = \frac{3 M^2_p}{8\pi} \left[H^2 - \frac{M^2_p}{12\pi} \left(
H'\right)^2\right].
\end{equation}
The equation (\ref{1}) can be simplified by introducing
the field $Q = e^{1/2\int\left[H - 2\ddot\phi_c/\phi_c\right]dt}
\Phi$
\begin{eqnarray}
\ddot Q &-& \frac{1}{a^2} \nabla^2 Q -
\left[
\frac{1}{4} \left( H
- 2 \frac{\ddot\phi_c}{\dot\phi_c}\right)^2
+\frac{1}{2}\left[
\dot H- 2\frac{d}{dt}\left(\frac{\ddot\phi_c}{ \dot\phi_c}\right)\right]
-2 \left(\dot H -H \frac{\ddot\phi_c}{\dot\phi_c}\right)
\right] Q
= 0\label{h}.
\end{eqnarray}
This field can be expanded in terms of the modes
$Q_k=e^{i\vec k.\vec x} \xi_k(t)$
\begin{equation}
Q(\vec x,t) = \frac{1}{(2\pi)^{3/2}} {\Large \int} d^3k \left[
\alpha_k Q_k(\vec x,t) + \alpha^{\dagger}_k Q^*_k(\vec x,t)\right],
\end{equation}
where $\alpha_k$ and $\alpha^{\dagger}_k$ are the annihilation
and creation operators that complies with the commutation relations
\begin{eqnarray}
\left[\alpha_k,\alpha^{\dagger}_{k'} \right]&=& \delta^{(3)}(k-k'), \label{c1}\\
\left[\alpha_k,\alpha_{k'} \right]&=&\left[
\alpha^{\dagger}_k,\alpha^{\dagger}_{k'} \right]=0.\label{c2}
\end{eqnarray}
The equation for the modes $Q_k$ is
\begin{equation}\label{xxi}
\ddot{Q}_k + \omega^2_k(t) \  Q_k =0,
\end{equation}
where $\omega^2_k = a^{-2}\left(k^2 - k^2_0\right)$ is the squared
time dependent frequency and $ k_0$ separates the infrared
and ultraviolet sectors, and is given by 
\begin{equation}
\frac{k^2_0}{a^2} =
\frac{1}{4} \left( H
- 2 \frac{\ddot\phi_c}{\dot\phi_c}\right)^2
+\frac{1}{2}\left[
\dot H- 2\frac{d}{dt}\left(\frac{\ddot\phi_c}{ \dot\phi_c}\right)\right]
-2 \left(\dot H -H \frac{\ddot\phi_c}{\dot\phi_c}\right)
\end{equation}
Since the field $Q$ satisfy a Klein-Gordon like equation on a
FRW background metric $\left<ds^2\right> = dt^2-a^2dx^2$, also satisfy
the commutation relationship
\begin{equation}
\left[Q(\vec x,t)\dot Q(\vec x',t)\right]
= i \delta^{(3)}\left(\vec x-\vec x'\right).
\end{equation}
This implies that the modes $Q_k$ are renormalized by the expression
\begin{equation}\label{ren}
\dot Q^*_k Q_k - \dot Q_k Q^*_k = i.
\end{equation}

\subsection{Particular solutions}

If the inflaton field oscillates around the minimum of the potential
at the end of inflation the particular solutions when $\dot\phi_c =0$
and $\ddot\phi_c=0$ are very important.

On the points $\dot\phi_c=0$ we obtain that $Q_k=0$. However, the
solutions for $\Phi_k$ are nonzero
\begin{equation}
\Phi_k = a^{-1} \phi^{0}_k.
\end{equation}
where $\phi^{0}_k$ is the initial amplitude for $\Phi_k$, for each
wavenumber $k$. This means that the amplitude of each mode
decreases exponentially with time.

Other interesting particular solution is located at the points
$\ddot\phi_c=0$, when the field is at the minumum of the potential.
In these points the equation (\ref{xxi}) adopts the form
\begin{equation}
\ddot Q_k
+  \left[ \frac{k^2}{a^2} - \left(\frac{H^2}{4} - \frac{3}{2} \dot H
\right)\right] Q_k =0,
\end{equation}
where $\Phi_k = a^{-1/2} Q_k$.

\subsection{The Sasaki-Mukhanov field}

A manner to study both, metric and
inflaton fluctuations, can be made by means of the
SM field\cite{sm}: $S=\phi + {\dot\phi_c \over H_c} \Phi$.
The modes of this field obeys the following equation
\begin{equation}\label{...}
\ddot S_k(t) + 3 H \dot S_k(t) + \left[ \frac{k^2}{a^2} +
V'' + 2 \frac{d}{dt} \left( \frac{\dot H}{H} + 3 H\right)\right] S_k(t)=0,
\end{equation}
where the modes $S_k$ complies with the renormalization condition
\begin{equation}
\dot S^*_k S_k - \dot S_k S^*_k = \frac{i}{a^3},
\end{equation}
so that $\left[S(\vec x,t), \dot S(\vec{x'},t)\right]={i\over a^3} \delta^{(3)}(
\vec x- \vec{x'})$.

\subsection{Power spectrum}

One can estimate the power spectrum of the fluctuations for the fields
$\Phi$ and $S$. The spectrum of the fluctuations for $\Phi$ are
\begin{equation}
{\cal P}_{\Phi} = \frac{k^3}{3\pi^2} \left|\Phi_k(t)\right|^2,
\end{equation}
whilst the power spectrum for the SM field is the same that
of the inflation field
\begin{equation}
{\cal P}_{S} = \frac{k^3}{2\pi^2} \left|S_k(t)\right|^2.
\end{equation}
It is well known from experimental data\cite{prl} that the universe
has a scale invariant power spectrum on cosmological scales.

\section{An Example:
Symmetric exponential $\phi_c$-potential: power-law inflation}

As a first example we consider a scalar potential
given by $V(\phi_c) = V_0 \  e^{2\alpha |\phi_c|}$, where
$\alpha^2={4\pi \over M^2_p p}$ gives the relationship
between $\alpha $ and the power of the expansion $p$.
This potential is related to a scale factor that evolves as $a \sim
t^p$ (with constant power $p$), which corresponds to a Hubble parameter
$H(t)=p/t$, which can be written in terms of the scalar field
\begin{equation}
H_c =\frac{\pi}{ M_p}
\left(\frac{32 V_0}{12\pi - \alpha^2 M^2_p}
\right)^{1/2} \  e^{\alpha |\phi_c|},
\end{equation}
where $V_0 = {3 M^2_p \over 8\pi} H^2_e \left[{12\pi - M^2_p \alpha^2
\over 12\pi}\right]$ and $H_e=p/t_e$ is the value of the Hubble
parameter at the end of inflation.
The temporal evolution for $|\phi_c(t)|$ is given by
\begin{equation}
|\phi_c(t)| = |\phi_0 | - \frac{1}{\alpha} {\rm ln}\left(\frac{t}{t_0}\right),
\end{equation}
where $t \geq t_0$. Since $\dot\phi_c = -{\rm sgn}(\phi_c){1\over \alpha t}$
and $\ddot\phi_c = {\rm sgn}(\phi_c){1\over \alpha t^2}$
(we assume ${\rm sgn}(\phi_c)= \pm 1$ for $\phi_c$ positive and
negative, respectively), the
equation that describes the evolution for $\Phi$ results
\begin{equation}
\ddot\Phi + \frac{(p+2)}{t} \dot\Phi - \frac{1}{a^2} \nabla^2 \Phi =0.
\end{equation}
After make the transformation $Q = \Phi e^{\int (p+2)t^{-1} dt}$,
we obtain the differential equation for $Q$
\begin{equation}
\ddot Q - \frac{1}{a^2} \nabla^2 Q - \left[\frac{p}{2}\left(\frac{p}{2}+1
\right)t^{-2} \right]Q=0.
\end{equation}
The general solution for the modes $Q_k(t)$ is
\begin{equation}
Q_k(t) = C_1 \sqrt{\frac{t}{t_0}} {\cal H}^{(1)}_{\nu_1}[x(t)]
+ C_2 \sqrt{\frac{t}{t_0}} {\cal H}^{(2)}_{\nu_1}[x(t)],
\end{equation}
where ($C_1$,$C_2$) are constants, (${\cal H}^{(1)}_{\nu_1}[x]$,
$H^{(2)}_{\nu_1}[x]$) are the Hankel functions of (first, second) kind
with $x(t) = {t^p_o k \over a_0 (p-1) t^{p-1}}$ and $\nu_1 = {p+1 \over
2(p-1)}$. Using the renormalization condition $\dot Q^*_k Q_k - \dot Q_k
Q^*_k = i$,
we obtain the Bunch-Davis vacuum\cite{BD} solution
($C_1=0$,$C_2 = \sqrt{\frac{\pi}{2(p-1)}}$) 
\begin{equation}\label{H}
Q_k(t) = \sqrt{\frac{\pi}{2}} \sqrt{\frac{t}{t_0(p-1)}}
{\cal H}^{(2)}_{\nu_1}[x(t)],
\end{equation}
In the UV sector the function ${\cal H}^{(2)}_{\nu_1}[x]$
adopts the
asymptotic expression (i.e., for $x \gg 1$)
\begin{equation}
{\cal H}^{(2)}_{\nu_1}[x] \simeq \sqrt{\frac{2}{\pi x}}\left[
{\rm cos}\left(x-\nu_1\pi/2 - \pi/4\right)-i \  {\rm sin}
\left(x-\nu_1\pi/2 - \pi/4\right)\right],
\end{equation}
whilst on the IR sector (i.e., for $x \ll 1$)
it tends asymptotically to
\begin{equation}
{\cal H}^{(2)}_{\nu_1}[x] \simeq \frac{1}{\Gamma(\nu_1+1)} \left(
\frac{x}{2}\right)^{\nu_1} - \frac{i}{\pi} \Gamma(\nu_1) \left(
\frac{x}{2}\right)^{-\nu_1}.
\end{equation}
The $\Phi$-squared field fluctuations on the IR sector are
$\left(\left<\Phi^2\right>\right)_{IR} =
{1 \over 2\pi^2} {\Large\int}^{\epsilon k_0(t)}_{0} dk k^2
\left|\Phi_k\right|^2$, and becomes
\begin{eqnarray}
\left(\left<\Phi^2\right>\right)_{IR} & \simeq &
\frac{1}{4}\left\{\frac{\left[\frac{t^p_0}{2 a_0(p-1)}\right]^{2\nu_1}
\left[\frac{\epsilon a_0 \sqrt{\frac{p}{2}\left(\frac{p}{2}+1\right)}}{t^p_0}
\right]^{\frac{2(2p-1)}{p-1}}}{
\pi t_0 \Gamma^2\left(\frac{3p-1}{2(p-1)}\right)(3p-1)}\right. \nonumber \\
& + & \left.
\frac{\Gamma^2\left(\nu_1\right) \left[\frac{t^p_0}{
2 a_0(p-1)}\right]^{-2\nu_1} \left[\frac{\epsilon a_0 \sqrt{
\frac{p}{2}\left(\frac{p}{2}+1\right)}}{t^p_0}\right]^{\frac{2(p-2)}{p-1}}}{
\pi^3 t_0 (p-3)} \right\}  t^{3-2\nu_1},\label{AA}
\end{eqnarray}
where $\epsilon =k^{(IR)}_{max} /k_p \ll 1$ is a dimensionless constant,
$k^{(IR)}_{max} = k_0(t_*)$ at the moment $t_*$ when the horizon entry and
$k_p$ is the Planckian wavenumber (i.e., the scale we choose as a cut-off
of all the spectrum). The power spectrum on the IR sector is
$\left.{\cal P}_{\Phi}\right|_{IR} \sim k^{3-2\nu_1}$.
Note that  $\left(\left<\Phi^2\right>\right)_{IR}$
increases for $p >2$, so that to the IR squared $\Phi$-fluctuations
remain almost constant on cosmological scales
we need $p \simeq 2$. We find that
a power close to $p =2$ give us a scale invariant power
spectrum (i.e., with
$\nu_1 \simeq 3/2$ for $\left(\left<\Phi^2\right>\right)_{IR}$.
Furthermore, density fluctuations for matter energy density are given
by $\delta\rho/\rho = -2\Phi$, so that $\left<\delta\rho^2\right>^{1/2}/\left<
\rho\right> \sim \left<\Phi^2\right>^{1/2}$.

On the other hand, in the UV sector these fluctuations are given by
\begin{equation} \label{BB}
\left(\left<\Phi^2\right>\right)_{UV} \simeq
\frac{a_0}{4 t^{p+1}_0 \pi^2}
\left\{\frac{k^2_p}{t^2} -
\frac{a^2_0}{t^{2p}} \left[\frac{p}{2} \left(\frac{p}{2}+1\right)\right]\right\}
t^{3-2\nu_1}.
\end{equation}
The power spectrum in this sector go as $\left.{\cal P}_{\Phi}\right|_{UV}
\sim k^4$.
We observe from eq. (\ref{BB})
that $\left(\left<\Phi^2\right>\right)_{UV}$
increases during inflation for $p > 3$. 
From the results (\ref{AA}) and (\ref{BB}) we obtain that
$1 < p \le 2$, because a power-law $p>2$
could give a very inhomogeneous universe on cosmological scales.
Since $\left(\left<\Phi^2\right>\right)_{UV} \ge 0$,
we obtain the condition
\begin{equation}
k^2_p -
\frac{a^2_0}{t^{2(p-1)}} \left[\frac{p}{2} \left(\frac{p}{2}+1\right)\right]
\ge 0.
\end{equation}
If $a_0=H^{-1}_0$ ($H_0$ is the initial value of the Hubble parameter),
inflation should ends at $t=t_e$, where
\begin{equation}\label{uaa}
t_e \simeq \left[\frac{k_p H_0}{\sqrt{\frac{p}{2}\left(\frac{p}{2}+1\right)}}
\right]^{\frac{1}{p-1}}.
\end{equation}
For example, for $k_p H_0 = 10^{11} \  M_p$ and $p=2$, we
obtain $t_e \simeq 5.8 \  10^{10} \  M^{-1}_p$.

Now we can study the evolution of the SM field fluctuations
$\left<S^2\right>$. The eq. (\ref{...}) written explicitely
for the model we are studying is
\begin{equation}
\ddot S_k + 3 p t^{-1} \dot S_k + \left(k H_0\right)^2 \left(\frac{t_0}{
t}\right)^{2p} S_k =0.
\end{equation}
Note that the last term inside the brackets in eq. (\ref{...}) becomes null.
The general solution of eq. (\ref{uaa})
can be written in terms of the Hankel functions
\begin{equation}
S_k(t) = A \left(\frac{t}{t_0}\right)^{\frac{1}{2}(1-3p))}
{\cal H}^{(1)}_{\nu_2}\left[\frac{k H_0 t^p_0 t^{1-p}}{p-1}\right]
+ B \left(\frac{t}{t_0}\right)^{\frac{1}{2}(1-3p))}
{\cal H}^{(2)}_{\nu_2}\left[\frac{k H_0 t^p_0 t^{1-p}}{p-1}\right],
\end{equation}
where $\nu_2={(3p-1) \over 2(p-1)}$. If we adopt the
Buch-Davis vacuum\cite{BD}: $A=0$, $B=\sqrt{{\pi \over 2(p-1)}}$, we obtain
\begin{equation}
S_k(t) = \sqrt{\frac{\pi t^{1-3p}}{2(p-1)
t^{1-3p}_0}} \  {\cal H}^{(2)}_{\nu_2}\left[\frac{k H_e t^p_0
t^{1-p}}{p-1}\right].
\end{equation}
The power spectrums in the extreme sectors of the spectrum go as
\begin{eqnarray}
&& \left.{\cal P}_S\right|_{UV} \sim k^4, \\
&& \left.{\cal P}_S\right|_{IR} \sim k^{3-2\nu_2}.
\end{eqnarray}
Note that $ \left.{\cal P}_S\right|_{IR}$ is scale invariant for
$\nu_2 =3/2$, which corresponds with $p \rightarrow \infty $.
The squared $S$-fluctuations on both, the UV and IR sectors are
\begin{eqnarray}
\left(\left<S^2\right>\right)_{UV} & \simeq &
\frac{t^{-2} t_0}{4\pi^2 H^3_e}
\left[k^2_p H^2_e \left(\frac{t_0}{t}\right)^{2p} t^2
- \left(\frac{9}{4} p^2 -\frac{15}{2} p +2\right)
\right], \label{uva} \\
\left(\left<S^2\right>\right)_{IR} & \simeq &
\frac{\Gamma^2(\nu_2) H^{-3}_e \epsilon^{3-2\nu_2}
\left(\frac{9}{4}p^2 -\frac{15}{2}p +2\right)^{\frac{3-2\nu_2}{2}}}{
8\pi^3 (p-1)^{1-2\nu_2} (3-2\nu_2) t_0} t^{-2}.\label{ira}
\end{eqnarray}
Since $\left(\left<S^2\right>\right)_{UV} \ge0$ during inflation, from
the eq. (\ref{uva}) we obtain the condition
\begin{equation}
\left[k^2_p H^2_e
\left(\frac{t_0}{t}\right)^{2p} t^{2}
- \left(\frac{9}{4} p^2 -\frac{15}{2} p +2\right)
\right] \ge 0.
\end{equation}
From this condition we obtain the time for which inflation ends. Hence,
one obtains
\begin{equation}
t_e \simeq \left(\frac{k_p H_e}{\sqrt{\frac{9}{4}p^2 -\frac{15}{2} p +2}}
\right)^{\frac{1}{p-1}},
\end{equation}
where, since
we require ${9\over 4}p^2 -{15\over 2} p +2 \ge 0$, must to holds $p \ge 3.04$.
For example, for $p=4$ and $k_p H_e = 10^{11} \  M_p$,
one obtains $t_e \simeq 2.8 \  10^{3} \  M^{-1}_p$, which
is incompatible with the value obtained from the evolution for
$\Phi$.
On the other hand
$\left(\left<S^2\right>\right)_{IR}$ decreases as $t^{-2}$ independently
on the value of the power $p$.

\section{Final Comments}

In this paper we have studied
the evolution of the fluctuations in a symmetric $\phi_c$-exponential
potential which provides a power-law expansion
using
both, the gauge invariant field $\Phi$ and the Sasaki-Mukhanov field.
This last takes into account simultaneously, the inflaton and metric
fluctuations.
The results obtained from the evolution of 
$\left(\left<\Phi^2\right>\right)$ and $\left(\left<S^2\right>\right)$
are different in both treatments.
The reason can be explained from the fact that
the field
$S=\phi + {\dot\phi_c \over H} \Phi $ is not gauge-invariant and hence
do not describes correctly the fluctuations for $\phi$ and $\Phi$.
The fluctuations are well described by the field $\Phi$
which is gauge invariant
and predicts a scale invariant power
spectrum on the IR sector for $p \rightarrow 2$.
Note that we have not considered back-reaction effects
which are related to a second-order metric tensor fluctuations. This topic
was considered by Abramo and Nambu, who investigated a renormalization-group
method for an inflationary universe\cite{Abramo,Nambu}.
A different approach to describe the metric fluctuations was
considered more recently by Lyth and Wands\cite{lyth}
(see also\cite{moroi}), who suggested that
curvature perturbation could be generated by a light
scalar field named curvaton.\\

\vskip .2cm
\centerline{\bf{Acknowledgements}}
\vskip .2cm
MB acknowledges CONICET, AGENCIA 
and Universidad Nacional de Mar del Plata
for financial support.\\

\end{document}